# Detection of preventable fetal distress during labor from scanned cardiotocogram tracings using deep learning

Martin G. Frasch*[1]; Shadrian B. Strong*[1]; David Nilosek[1]; Joshua Leaverton[1]; Barry S. Schifrin[1]

*Equal contributions

[1] Heart Rate AI, Inc

**Address of correspondence:**
Martin G. Frasch
Department of Obstetrics and Gynecology
University of Washington
1959 NE Pacific St
Box 356460
Seattle, WA 98195
Phone: +1-206-543-5892
Fax: +1-206-543-3915
Email: mfrasch@uw.edu

**Abstract**

Despite broad application during labor and delivery, there remains considerable debate about the value of electronic fetal monitoring (EFM). EFM includes the surveillance of the fetal heart rate (FHR) patterns in conjunction with the mother's uterine contractions providing a wealth of data about fetal behavior and the threat of diminished oxygenation and perfusion. Adverse outcomes universally associate a fetal injury with the failure to timely respond to FHR pattern information. Historically, the EFM data, stored digitally, are available only as rasterized pdf images for contemporary or historical discussion and examination. In reality, however, they are rarely reviewed systematically. Using a unique archive of EFM collected over 50 years of practice in conjunction with adverse outcomes, we present a deep learning framework for training and detection of incipient or past fetal injury. We report 94% accuracy in identifying early, preventable fetal injury intrapartum. This framework is suited for automating an early warning and decision support system for maintaining fetal well-being during the stresses of labor. Ultimately, such a system could enable a physician to timely respond during labor and prevent adverse outcomes. When adverse outcomes cannot be avoided, they can provide guidance to the early neuroprotective treatment of the newborn.

**Introduction**

In the United States, there are approximately 4 million births per year. Over 85% of them are accompanied by electronic fetal monitoring (EFM) in labor to safeguard fetal/neonatal well-being. This surveillance of the FHR pattern (rhythm) in conjunction with the mother's uterine contractions, provides a wealth of data about fetal behavior and the threat of diminished oxygenation and perfusion. Fifty years after its introduction, however, fetal monitoring continues to inspire debate about its value and especially its role in increasing the cesarean section rate as well as being a "litogen" - a stimulus to allegations of medical malpractice.[1–9] Reviews of adverse labor outcomes in numerous countries universally associate adverse fetal outcomes with the failure to timely respond to the FHR pattern information.[10–12] Indeed, various sources affirm that misinterpretation of EFM (or the uncertainty with patterns) has contributed to the significantly increased use of Cesarean delivery from 5% in the 1970s to >30% today[13,14], leading to increased expenditures, costing the country[13,14] over $1 Billion per year per 5% of additional cesarean deliveries.[15] Obstetrical liability costs the country approximately $40 Billion per year of which 70% is accounted for by uncertainty about EFM interpretation and related brain injury upon delivery.[14,15]

Earlier and more precise detection of babies suffering preventable injury during labor is urgently needed. Prevention can be e.g., by an emergency delivery. Benefits include better maternal and child health thanks to a lower cesarean delivery rate and immediate neonatal monitoring of heart rate pattern, i.e., having the baby continuously monitored for at least 15 minutes after delivery. Here, babies seen to be at risk can be evaluated and more aggressively treated earlier than currently undertaken.

Historically, the EFM data, stored digitally, are available only as rasterized pdf images for contemporary or historical discussion, and examination (Figure 1). However, they are rarely reviewed systematically. In the case of a medical-legal review, it is the paper copy of the tracing, exclusively, that is likely available and consulted.

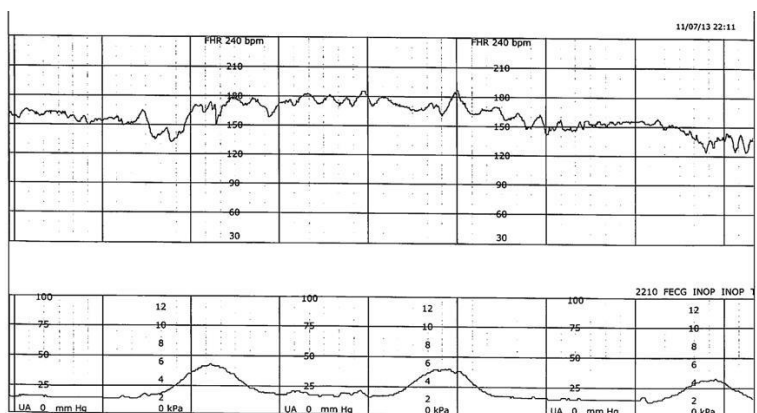

**Figure 1.** Example of FHR (top) and uterine contraction (bottom) during labor, captured simultaneously and stored electronically in digital format but available only as a rasterized pdf document.

We propose a deep learning-based approach to this challenge. It is based on a unique archive collected over four decades of EFM documents of babies with known adverse outcomes. They provide many unique examples of the broad range of healthy, threatened, and injured fetuses along with their long-term outcomes. Consequently, this archive is ideal for automating an early warning (guidance) system for maintaining fetal well-being during the stresses of labor and delivery that could ultimately enable a health care provider to timely respond during labor and prevent adverse outcomes. When adverse outcomes cannot be avoided, they guide the early neuroprotective treatment of the newborn. This system utilizes a unique classification of heart rate and contraction patterns (details in Methods), including specific identifiable indicators ('point A' and 'point B') of the need for attention by the provider.[16–19]

**Methods**

### *1. Data*

For this pilot study, a convenience sample of 36 tracings was selected. All tracings derived from singleton pregnancies at term undergoing a trial of labor with a fetal monitor in place as previously described.[17] Each tracing was considered normal at the onset of monitoring. The majority of features derived from conventional guidelines (ACOG) including baseline rate, variability, accelerations, and decelerations. For this study, however, certain operational definitions of heart rate patterns (Table 1) and uterine contractions (Table 2) were modified. These included the basal rate, the use of relative bradycardia and tachycardia, and the pattern of recovery of the deceleration.

### *2. Identification of EFM features*

**Table 1. Definitions of EFM patterns**

| | |
|---|---|
| **Basal heart rate** | The baseline FHR established at the beginning of labor with fetus quiescent. |
| **Tachycardia** | Absolute – sustained (>10 min) baseline heart rate above 155 bpm<br>Relative – sustained (>10 min) baseline heart rate >15 above basal rate |
| **Bradycardia** | Absolute – sustained (>10 min) baseline heart rate below 110 bpm<br>Relative – sustained (>10 min) baseline heart rate >15 bpm below basal rate |
| **Deceleration recovery** | The response of the fetus to a deceleration |
| **Normal response** | Return to the previously normal baseline rate and variability |
| **Adverse response** | Applies to the recovery of the deceleration, but may persist as a feature of the subsequent baseline heart rate |
| **Overshoot** | An acceleration of the FHR immediately following a deceleration with duration proportional to the amplitude of the preceding deceleration. Usually associated with alterations in baseline rate and variability. |
| **Delayed return** | A "slow return" to the baseline - likely a sustained elevation of fetal BP in anticipation of recovery. |
| **Peaked return** | An abrupt peak at the end of a deceleration followed by a late deceleration. An ominous commentary usually leading to fetal death. |
| **Decreased /absent variability** | Persistent diminution in baseline variability <6 bpm |
| **Increased variability** | Persistent or transient elevation of variability >25 bpm |
| **Sinusoidal pattern** | Visually apparent, smooth, sine wave-like undulating pattern in FHR baseline with a cycle frequency of 3–5 per minute. Occurs in the absence of normal CTG pattern nearby. May be brief or persistent. |
| **Checkmark pattern** | Unique pattern seen in neurologically compromised / asphyxiated fetus suggesting repetitive "checkmarks" ( ) of varying duration – frequently elicited by a preceding deceleration. |
| **Sawtooth pattern** | Rapid, high frequency (20+cpm), low amplitude (<15 bpm), peaked oscillations in the heart rate that generally increase in frequency and decrease in amplitude over time |

| | | |
|---|---|---|
| **Conversion** | An CTG pattern in which there is a dramatic change in rate, variability and pattern of deceleration within 1 to 2 contractions. – suggests fetal ischemic injury. | |

**Table 2. Definition of excessive uterine activity**

| Contraction Parameter | Average | Excessive |
|---|---|---|
| **Frequency** | 2 – 4.5 UC / 10 minutes | >5 / 10 minutes (x2) |
| **Intensity** | 25 – 75 mmHg | Not defined! |
| **Duration** | 60 – 90 sec. | >90 sec. |
| **Resting Tone** | 12 – 20 mmHg | >20 mmHg |
| **Interval between peaks** | 2 – 4 minutes | <120 sec |
| **Rest time*** | 50 – 75% | < 50% |
| **Montevideo Units -** | DO NOT USE! | |

* Rest time - interval when contractions and pushing are absent.

Abbreviations: UC, uterine contractions = millimeters of mercury

The TRACING is defined at the outset of monitoring as NORMAL or ABNORMAL.

A NORMAL tracing is characterized by a stable baseline heart rate between 110-155 bpm, with moderate variability and absent decelerations. An ABNORMAL tracing is characterized by at least one of the following features:

- Baseline heart rate: <100, >155, arrhythmia, unstable / indeterminate
- Baseline variability: Absent, decreased (<6bpm), Increased (>25bpm)
- Decelerations: Late, variable, undefined.

Thus, for NORMAL DECELERATIONS no immediate action is required. They return promptly to the previously normal baseline variability (5-15 bpm peak to trough and chaotic, pseudorandom), and heart rate (usually 110-155 bpm and stable); each fetus has an individually unique baseline. A "normal" deceleration returns to baseline without changing trajectory and upon reaching the previous baseline rate remains there. These features pertain regardless of amplitude, duration, and timing of the deceleration and signify the comfortable compensation for the alteration in blood flow represented by the contraction.

**POINT A**

Point A denotes the time when the recovery of the deceleration is no longer "normal" and those additional compensatory activities are invoked by the fetus to maintain homeostasis. The detection of Point A signifies that increased attention and conservative measures are needed to restore homeostasis to the previously normal tracing. These features include:

> A. Delayed return to baseline: includes a change in the trajectory of the recovery such that the return to baseline is delayed beyond the end of the contraction.
> B. Period of increased variability: peak to trough greater than 20 bpm, frequency 5 to 10 cycles per minute. Duration is also influenced by the appearance of a subsequent contraction during which time the pattern disappears – taken over by the deceleration.
> C. Overshoot: An acceleration following the upslope of the return of the deceleration lasting 15 seconds or more prior to the return to the baseline.
> D. Transient (usually at least 1 minute) return to a higher baseline by at least 15 bpm, duration affected by next contraction, compared to the previously stable baseline.
> E. Transient (at least 1 minute) return to a lower baseline by at least 10 bpm compared to the previously normal baseline.

The detection of Point A alerts the health care provider to the need for at least conservative intervention regarding the maternal condition, the frequency of contractions or expulsive efforts during the 2nd stage of labor. Point A is identified sooner if an excessive uterine activity is present.

**Point B**

Point B represents the attempt to define significant fetal compromise or injury, irrespective of the perceived amount of acidosis (pH) in the fetus. No clinical circumstances were used in the definition of Point B. Point B was identified by the subject expert (BSS) via a custom-created digital interface (AWS) allowing us to feed the annotations directly into the deep learning model.

These features include:

- Sustained return to a baseline with diminished/absent baseline variability, usually accompanied by a rise in the baseline heart rate.
- This sustained change in baseline rate and variability with adverse features (Table 1) occurring within 5 to 10 minutes of a previously normal rate and variability - usually with recurrent variable decelerations.

### 3. Deep Learning Pipeline

We present a method for automated extraction of features in FHR and uterine contractions (UC) which are outlined in the above section.

Briefly, the method includes acquiring a set of non-digitized charts, digitally assigning markers to predetermined features in the charts, supplying the assigned marker-feature sets to a supervised model, statistically iterating over the assigned sets, automatically assessing model performance, and applying the model to new sets of charts to extract non-assigned predetermined features.

A method for automated chart processing includes analyzing time-series sets of non-digitized charts of FHR and/or concurrent maternal UC to digitally associate markers with fetal signatures, using the associated groups for supervised training of an artificial intelligence model, determining accuracy and precision of the model, and applying the trained model to automatically process new time-series sets of one or more charts of FHR and concurrent maternal UC, having one or more un-associated fetal signatures.

More specifically, we treat scanned EFM recordings as non-vectorized images, similar to digital photographs, and apply supervised machine learning to extract and process features to train an artificial intelligence model. An image is supplied to a convolution neural network (CNN) model. The image is represented as one or more numeric arrays of pixel values with varying signal counts associated with the pixel content. The pixel content is dictated by the amount of red, green, blue, or other spectral bands that the pixel may receive, and is an integer number in one or more dimensions. The CNN is represented as a set of algorithmic layers into which the numeric pixel data are sent. It consists of a series of convolutional layers, nonlinear layers, pooling layers, and fully connected layers. Each such layer may be considered an individual set of equations, where the output of one equation becomes the input to another. The CNN eliminates the need for manual feature extraction, as the features are acquired through the passing of the pixel data to one or more other layers, and correlations are extracted and weighted as a consequence of the layer transitions.

For a supervised approach, the model is provided with one or more labeled examples of the important features (labels, markers, and/or training data) for the CNN to generate correlations among the images with the features. The CNN then adapts its algorithm to try to predict the relevant features through statistical iteration of the features extracted from one or more layers in the CNN.

We implement a Single Shot Detector (SSD) algorithm to achieve this goal.[20] It utilizes a standard CNN network (e.g., VGG-16) with an additional set of convolution layers to identify discrete locations of one or more features in one or more images.

Through a single pass in the CNN, the weighted correlations meant to describe the relevant features are tested against ground truth data (validation data), separate from training data. The goal of this statistically iterative operation is to minimize a loss function between the predicted correlations and the truth values through adaptively updating the weights of the predicted function. The process of adjusting the weights continues until a minimum statistical loss is obtained.

The output model and weights are then used for inference against the withheld (unseen) data set to extract similar relevant information.

In this study, we implemented a conventional 80/20 train/test split. This corresponded to 26.4 hours of training on EFM image information and 6.6 hours used for testing (validation).
That is, the EFM images were flipped/translated and the noise was added to represent more of the variability observed in the original pdfs.

The codebase is available here: https://github.com/zhreshold/mxnet-ssd

**Results**

The demographics, clinical characteristics are summarized in Table 3. There were eleven outcomes with pH < 7.00. Table 3 also denotes the duration and timing of the first and second stage of pushing and Point A.

**Table 3. Cohort characteristics.**

|  | Age, years | EGA, weeks | BMI | BWT, g | Apgar 1 | Apgar 5 |
|---|---|---|---|---|---|---|
| Median (25th;75th percentile) | 27 (21;30) | 40 (39;40) | 31 (27;36) | 3325 (3070;3602) | 2 (1;4) | 6 (4;7) |

|  | Temporal characteristics of labor (hours:minutes) | | | | |
|---|---|---|---|---|---|
|  | 1st stage | 2nd stage | Labor | Point A | Point A to Delivery |
| Median (25th;75th percentile) | 18:36 (15:04;26:34) | 2:45 (1:41;3:59) | 24:50 (16:41;48:00) | 13:38 (5:44;20:43) | 4:01 (1:52;5:18) |

As a step towards developing this proactive fetal surveillance system, we have created an artificial intelligence model using a basic SSD deep learning approach to retrospectively identify critical features in the EFM data (cardiotocography) from the rasterized pdf directly (Figure 2). This model creates a classification of the pattern and identifies critical features of the tracing that indicate critical and timely points of either conservative or operative intervention, 'Point A' and 'Point B'. Here, in the initial implementation, we focused on predicting 'Point A'.

This novel application of using pdf rasterized plots as an image detection deep learning problem facilitates (1) quick and efficient deployment against a large record of data without chart digitizing and (2) packaging and deployment as a light-weight or mobilenet [21] application useful for immediate integration with a mobile device, post-event.

The model achieved an accuracy of 93.6% in identifying Point A against a small dataset with limited variability in features.

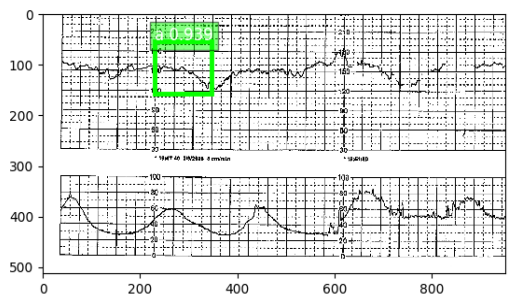 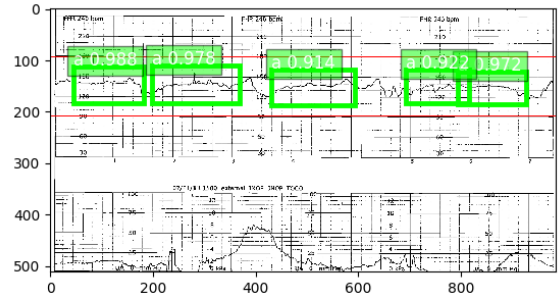

**Figure 2**. (Left) 'Point A' identified with accuracy 93.6% using an SSD trained on the pdf chart images. (Right) Numerous occurrences of 'Point A' with high confidence in green. Red indicates the truth 'Point A' duration.

**Discussion**

We successfully implemented automated identification of Point A indicating a preventable fetal injury which is of highest relevance for real-time clinical implementation of such an algorithm. Based on our observations of ~5,000 cases with brain injury as birth outcome, 20% of normal patients reach Point A. About 25% of these revert to normal. Point B is reached in about 0.5% of the population and about 30-40% of our observations. We leave it to future work to implement the prediction of Point B, as this will require training on larger datasets. These points, together with other key signs in the FHR can be displayed for the obstetrical care provider as part of an early alert and decision support system. Consequently, the visual signature for training the SSD is extracted similarly to the method utilized by the physician. The time-series nature of the FHR may be exploited with an additional application of a Long Short-Term Memory (LSTM)[22] model for consistent identification and tracking as a function of event duration. However, to date, only the SSD has been deployed.

It is important to emphasize that the training of the model was not based on the detection of acidosis or even low Apgar score, but whether or not conservative intervention based on heart rate pattern (Point A) was deemed necessary and whether criteria were met for the presumptive diagnosis of fetal neurological injury (Point B) as described previously.[16] There was no attempt to correlate the outcome results with either pH or Apgar score of the newborn.

Note the relatively early appearance of Point A during late 1st or early 2nd stage of pushing. Consequently, the detection of Point A leaves time for preventive intervention and can help avoid emergency delivery and lasting injury.

In response to Point A, conservative rehabilitative measures include:

- Diminishing the frequency of uterine contractions;
- Diminishing/ceasing pushing during the 2nd stage of labor;
- Decreasing infusion of oxytocin;
- Assessing the feasibility of safe vaginal delivery.

The objective of our study was to use deep learning as a preventive intervention rather than one of immediate intervention ("rescue") by identifying the point in the tracing, before fetal acidosis has developed and where conservative measures can be expected to restore the tracing to normal by the implementation of conservative measures. We see no benefit in employing an artificial intelligence system to define acidosis and, simultaneously, the need to rescue the fetus that may have already become injured. The system is designed to work with fetuses with initially normal tracings as no real benefit can be calculated from any algorithm that begins with an abnormal tracing where the options for prevention are limited and early delivery is likely.[23]

The presented approach of presenting and interpreting existing clinical data and annotating the EFM record during labor can dramatically reduce the need for urgent deliveries and significantly improve the outcomes of babies and mothers in labor and for the neonate in the nursery. Improved outcomes, less urgency, fewer rescues, and better documentation could be a game-changer for the care of pregnant women and children and the defense of allegations of obstetrical malpractice.

In future work, to boost the present performance results, alternatively or additionally, RCNN, LSTM, RNN, support vector machine, random forest, instance segmentation, image classification techniques, and/or other deep learning algorithms and/or other machine learning techniques can be applied.

The new EFM data can be supplied to the trained model in a format different from the format of the original training/testing images. For example, the EFM data can be supplied in the format of digitized charts, tabularly represented data, a signal received from one or more devices, etc. In other words, once the model has been trained, it can be configured to work on similar features provided in the same and/or other data formats, including live data. Such an approach allows the model to identify one or more features of interest, and also the location of those features in the chart(s). This location can be correlated with a time and/or other dependent variables within the chart and/or a set of charts.

These features of our approach make it attractive to the electronic medical record and physiological monitoring applications well beyond EFM.

We have shown the feasibility of a DL approach to scan and detect the ability of the fetus to handle the trial of labor using standard fetal heart rate and uterine activity chart tracings presented to artificial intelligence in the form of images, the format in which most of such tracings are still stored and presented to the experts for the determination of the need for intervention and the timing of the fetal injury. Our DL approach detects these factors with over 90% accuracy (compared to expert scoring).